\documentclass[useAMS,usenatbib]{mn2e}
\usepackage{graphicx}
\usepackage{dcolumn}
\usepackage{bm}

\usepackage{amsmath}
\usepackage{amssymb}

\title[Relic signal from dark matter particles annihilation
]{Dark matter annihilation at cosmological redshifts: possible
relic signal from weakly interacting massive particles
 annihilation}

\author[A. N. Baushev]{A. N. Baushev$^{1}$\\
$^{1}$Bogoliubov Laboratory of Theoretical Physics, Joint Institute for Nuclear Research\\
141980 Dubna, Moscow Region, Russia}
\begin{document}

\date{}

\pagerange{\pageref{firstpage}--\pageref{lastpage}} \pubyear{2009}

\maketitle

\label{firstpage}

\begin{abstract}
We discuss the possibility to observe  the products of dark matter annihilation
that was going on in the early Universe. Of all the particles that could be generated by this
process we consider only photons, as they are both uncharged and easily detectable. The earlier
the Universe was, the higher the dark matter concentration $n$ and the annihilation rate
(proportional to $n^2$) were. However, the emission from the very early Universe cannot reach us because
of the opacity. The main part of the signal was generated at the moment the Universe had just
become transparent for the photons produced by the annihilation. Thus, the dark matter
annihilation in the early Universe should have created a sort of relic emission. We obtain its flux
and the spectrum.

If weakly interacting massive particles (WIMPs) constitute dark matter, it is
shown that we may expect  an extragalactic gamma-ray signal in the energy range
0.5 - 20~{MeV} with a maximum near 8~{MeV}. We show that an experimentally
observed excess in the gamma-ray background at 0.5 - 20~{MeV} could be created
by the relic WIMPs annihilation only if the dark matter structures in the
universe had appeared before the universe became transparent for the
annihilation products ($z \simeq 300$). We discuss in more detail physical
conditions whereby this interpretation could be possible.
\end{abstract}


\section{Introduction}
Though the cosmological measurements \citep{cosmol} show that there must be
approximately five times as much dark matter as all baryon one, its physical
nature remains unknown. Now the most commonly used hypothesis is that it
consists of some elementary particles generated in the early Universe
(hereafter we will call them Dark Matter Particles, DMPs). These particles are
uncharged and do not interact strongly; there are telling arguments to believe
that they were cold ($\upsilon\ll c$) in the epoch when the relic radiation was
generated. It is worth mentioning that a particle with suitable properties
hasn't been discovered yet, in spite of no lack of theoretical candidates
predicted by various quantum field theory models.

If the premise is true, the dark matter is a mixture of equal quantities of
particles and antiparticles, and they must collide and annihilate wherever the
dark matter is present. Experimental observation of such a process would give
us some valuable information about the DMP nature. In this article, we consider
the dark matter annihilation in the epoch near the relic radiation formation
($z\sim 1000$). At that time, the average dark matter particle concentration
$n\propto (z+1)^3$ was nine orders higher than now and four orders higher than
in our Galaxy near the Sun system. So we may expect that the annihilation, the
rate of which is proportional to $n^2$, was very intensive in that epoch.

Of all the particles that can be generated by the dark matter annihilation we
will consider only photons, as they are both uncharged and easily detectable.
Uncharged particles do not interact with the magnetic field of the Galaxy,
which allows one to measure the extragalactic background reliably enough.

The earlier the Universe was, the higher the DM density and the annihilation
rate were. However, the emission from the very early Universe cannot reach us
because of the opacity. The main part of the signal was generated at the moment
the Universe had just become transparent for the photons produced by the
annihilation. Later the dark matter density rapidly dropped, decreasing the
signal. The moment (and its redshift) depends, of course, on the characteristic
energy of the photons, in other words, on the nature of the dark matter.

Thus, the DM annihilation in the early Universe should have produced a sort of
relic emission. We obtain its flux and the spectrum. On the one hand, such an
emission can be detected in the spectrum of the extragalactic background. A
distinguishing feature of this radiation should be its high isotropy. On the
other hand, the absence of such a signal can impose a severe limitation on a
dark matter model.

In calculations we assume that the dark matter is homogeneous. This is the
simplest, but by no means the most natural supposition. Possible amplitude of
the dark matter density perturbations and their influence on the annihilation
signal will be discussed at the end of the article.

In the second part of the article, the case of the most popular dark matter
candidate - Weakly Interacting Massive Particles (WIMPs) - is considered in
more detail. We demonstrate that the relic signal from the WIMPs annihilation
might have been already observed.

\section{Calculations}

The metric of a homogeneous isotropic universe can be represented
as $ds^2=c^2dt^2-a^2(t) dl^2$, where $dl$ is an element of
three-dimensional length \citep{teorpol}. In agreement with the
observations \citep{gorbrub} we assume zero three-dimensional
curvature of our Universe. We introduce polar coordinates with the
centre at the local observer. Then the metric can be written as:
\begin{equation}
ds^2=c^2dt^2 - a^{2}(t)\,[\,dr^2 + r^2 (d\zeta^2+\sin^2\! \zeta\; d\xi^2)\,] \label{a1}
\end{equation}
We choose the normalization of $a$ so that the contemporary value $a_0=1$
(hereafter the subscript $0$ is used to describe the present-day values of
quantities). Then $z$ and $a$ are related by the following equation:
\begin{equation}
z+1=\dfrac{1}{a} \label{b1}
\end{equation}
The number of annihilations in the volume $dV$ in an interval of the proper
time $d\tau$ is (we should remind once again that we consider a homogeneous
dark matter distribution)
\begin{equation}
\dfrac12 \langle\sigma\upsilon\rangle\: n^2 d\tau \, dV  \label{a4}
\end{equation}
Here the multiplier $\frac12$ takes into account that the annihilation is only
possible if a particle collides with an antiparticle\footnote{We do not
consider the situation when the dark matter particle is identical to its
antiparticle. In this case the multiplier should be $1$ instead of $\frac12$}.

Here we should make two temporary simplifying suppositions: we assume that one
act of annihilation produces one photon of fixed energy $\beta$ and that the
Universe is transparent for the photons (later we will generalize the result
taking into account opacity and arbitrary spectrum of the emitting photons).
Let us consider a three-dimensional space element that is part of a spherical
layer $[r; r+dr]$ viewed by the local observer from a solid angle $do$. If the
effective area of the detector is $dS$, the number of photons arriving at it in
an interval of time $dt$ is:
\begin{equation}
P=\dfrac{\langle\sigma\upsilon\rangle}{8\pi}\: \dfrac{n_0^2}{a^2}\: dt \, do\, dr\, dS  \label{a7}
\end{equation}
Photons emitted with the fixed energy $\beta$ arrive at the observer with a smaller energy
$\varepsilon$ in consequence of the redshift; $\beta$ and $\varepsilon$ are related by the usual
law
\begin{equation}
\dfrac{\varepsilon}{\beta}=a  \label{a8}
\end{equation}
where $a$ corresponds to the moment of emission. The bigger $r$ is, the smaller $a$ was.  In order
to find the relationship between them, we should write the equation of motion of a photon, that is $ds=0$.
From (\ref{a1}) we obtain for the radial motion $c\, dt=a\, dr$; whence it follows:
\begin{equation}
dr=\dfrac{c}{a}\, dt= \dfrac{c}{a} \dfrac{dt}{da}\, da= \dfrac{c}{a \dot a} \, da\label{a9}
\end{equation}
The expansion of the Universe after the radiation-dominated epoch
can be described as \citep{gorbrub}:
\begin{equation}
\dfrac{\dot a}{a}= H_0 \sqrt{\Omega_m \left(\dfrac{1}{a}\right) ^3+\Omega_\Lambda} \label{a10}
\end{equation}
As we will see, the main part of emission appears when $\frac{1}{a}\gg 10$, and we may neglect
$\Omega_\Lambda$ in this formula. Thus,
\begin{equation}
\dot a= \dfrac{H_0 \sqrt{\Omega_m}}{\sqrt{a}}  \label{a11}
\end{equation}
Substituting (\ref{a9}) and (\ref{a11}) into (\ref{a7}) we get:
\begin{equation}
P=\dfrac{\langle\sigma\upsilon\rangle}{8\pi H_0 \sqrt{\Omega_m}}\: \dfrac{n_0^2 c}{a^2 \sqrt{a}}\:
dt \, do\, da\, dS \label{a12}
\end{equation}
or, if we replace $a$ with $\varepsilon/\beta$, in accordance with (\ref{a8}),
\begin{eqnarray}
P=\dfrac{\langle\sigma\upsilon\rangle\, n_0^2 c}{8\pi H_0 \sqrt{\Omega_m}}\:
\left(\dfrac{\beta}{\varepsilon}\right)^{\displaystyle{\frac{5}{2}}} \: dt \, do\,
d\frac{\varepsilon}{\beta}\, dS = \nonumber \\
=\dfrac{\langle\sigma\upsilon\rangle\, n_0^2 c}{8\pi H_0 \sqrt{\Omega_m}}\:
\dfrac{\beta\sqrt{\beta}}{\varepsilon^2 \sqrt{\varepsilon}} \: dt \, do\, d\varepsilon\, dS
\label{a13}
\end{eqnarray}
So we obtain the sought-for spectral intensity $Q$ of the photon flux, i.e., the number of photons
that come to the local observer from unit solid angle per unit time per unit of area per unit
energy interval is:
\begin{equation}
Q=\dfrac{\langle\sigma\upsilon\rangle\, n_0^2 c}{8\pi H_0 \sqrt{\Omega_m}}\:
\dfrac{\beta\sqrt{\beta}}{\varepsilon^2 \sqrt{\varepsilon}} \label{a14}
\end{equation}
One can see that the integral of this equation over the energy diverges when $\varepsilon\to 0$. Of
course, to accomplish this, arbitrary early Universe should be transparent for the photons, which
 is not the case. The photons interact with the baryonic matter; let us assume that this process can be
characterized by some averaged cross-section $\aleph$. Then the number of interactions of the
primary photons per an element of physical length $dl=a\, dr$ is
\begin{equation}
\dfrac{dN}{N}=-\aleph n^b dl =-\aleph \dfrac{n^b_0}{a^3}\, a dr= -\aleph c n^b_0\, \dfrac{da}{a^3
\dot a} \label{a15}
\end{equation}
Here $n^b$ is the baryon concentration $n^b=n^b_0/a^3$; we also
used (\ref{a9}). In order to integrate this equation, we suppose
that $\aleph$ is constant. This assumption is usually not very
rigorous. Actually, it means that we neglect the cross-section
dependence on the photon energy (which strongly changes during
its propagation via the redshift) and possible phase transitions
in the Universe. However, since the baryon concentration $n^b$
rapidly decreases with the Universe expansion, only the moments
just after the emission contribute noticeably to the opacity. So
for the constant cross-section approximation to be feasible we
need only that the dependence of the cross-section on the
energy is not very abrupt, which is usually the case.  The only
phase transition which occurs in the Universe near $z\sim 1000$
is recombination. It changes drastically the cross-section of
photons with an energy $\lesssim 10$~{eV}, the approximation
$\aleph=\it{const}$ is obviously wrong in this case. However, the
energy of photons generated by the dark matter annihilation is
most likely considerably higher, and the recombination does not
affect much their cross-section.

Here we also suppose that the photon interaction with the baryonic
matter is pure absorption and interacting photons just disappear.
This assumption is more rough: the contribution of scattering can
be significant (see \citet{rassey} for details), and we
underestimate the soft part of the spectrum. However, the main
characteristics of the spectrum (the characteristic energy and the
intensity) are determined by the moment the Universe becomes
transparent for the photons produced by the annihilation, and this
moment does not depend much on whether it is absorption or
scattering. Consequently, the assumption is quite acceptable for
an estimation model. Below we will improve it.

Moreover, we introduce a new variable $\chi$ for the scaling factor of the Universe. It has exactly
the same meaning as $a$, but $a$ represents the scaling factor at the moment when the photon is
emitted, while $\chi$ represents the scaling factor changing during the photon propagation to the
observer. Consequently, $\chi$ changes from $a$ to $a_0\equiv 1$. Then (\ref{a15}) can be rewritten
as
\begin{equation}
\dfrac{dN}{N}= -\aleph c n^b_0\, \dfrac{d\chi}{\chi^3 \dot \chi}= \dfrac{-\aleph c
n^b_0\,d\chi}{\chi^2 \sqrt{\chi}\, H_0  \sqrt{\Omega_m+\Omega_\Lambda \chi^3}} \label{b2}
\end{equation}
We used (\ref{a10}) to obtain it. Since the right part of this equation rapidly drops with $\chi$
growing, we can simplify it:
\begin{equation}
\dfrac{dN}{N}=  -\aleph c n^b_0\, \dfrac{d\chi}{\chi^2 \sqrt{\chi}\, H_0  \sqrt{\Omega_m}}
\label{b3}
\end{equation}

Now equation (\ref{a15}) can be easily integrated.
\begin{equation}
\dfrac{N}{N_{initial}}=\exp\left(-\aleph c n^b_0\,\int_a^{\chi}\!\dfrac{d\chi}{\chi^2 \sqrt{\chi}\,
H_0  \sqrt{\Omega_m}}\right) \label{a16}
\end{equation}
It is convenient to introduce a new constant
\begin{equation}
\wp= \dfrac{\aleph c n^b_0}{H_0 \sqrt{\Omega_m}} \label{b4}
\end{equation}
From (\ref{a16}) we obtain:
\begin{equation}
\frac{N}{N_{initial}}=\exp\left(-\frac{2}{3} \dfrac{\wp}{a\sqrt{a}} + \frac{2}{3}
\frac{\wp}{\chi\sqrt{\chi}}\right)\label{a17}
\end{equation}
As it has been already mentioned, $\chi$ changes from $a$ to $a_0=1$. Since $a\ll a_0=1$, we may
omit the last term of (\ref{a17}). Replacing $a$ with $\varepsilon/\beta$ according to (\ref{a8})
we obtain
\begin{equation}
\dfrac{N}{N_{initial}}=\exp\left(-\frac{2}{3} \wp \dfrac{\beta \sqrt{\beta}}{\varepsilon
\sqrt{\varepsilon}}\right) \label{a19}
\end{equation}
In order to allow for the opacity of the early Universe, we should multiply equation (\ref{a14}) by
this exponential factor. Finally, we obtain:
\begin{equation}
Q=\dfrac{\langle\sigma\upsilon\rangle\, n_0^2 c}{8\pi H_0 \sqrt{\Omega_m}}\:
\dfrac{\beta\sqrt{\beta}}{\varepsilon^2 \sqrt{\varepsilon}}\: \exp\!\left(-\frac23 \wp \dfrac{\beta
\sqrt{\beta}}{\varepsilon \sqrt{\varepsilon}}\right) \label{a20}
\end{equation}
Since we supposed that one act of annihilation produces one photon of fixed energy $\beta$, this
equation actually specifies only a Green's function. If the produced photons have some distribution
$f(\beta)\, d\beta$, we should convolute $Q(\varepsilon,\beta)$, defined by (\ref{a20}), with the
distribution $f$ to obtain the spectrum that appears to the viewer:
\begin{equation}
\tilde Q(\varepsilon)=\int Q(\varepsilon,\beta) f(\beta)\, d\beta  \label{a21}
\end{equation}
Nevertheless, the spectrum given by (\ref{a20}) is fairly wide, and provided that the distribution
$f(\beta)$ is not too broad the shape of the resulting spectrum remains similar to (\ref{a20}).

\section{The case of WIMP dark matter}
Equations (\ref{a20}), (\ref{a21}) are applicable in quite general
cases, but they were obtained on a rather rough assumption that
the photon interaction with the baryonic matter is pure
absorption. In order to make the model more realistic, we should
consider a certain DMP candidate. If we make a very natural
supposition that the DMPs were in thermal equilibrium with other
particles in the early Universe, we can estimate their
annihilation cross section \citep{sechen, kazakov}:
\begin{equation}
\langle\sigma\upsilon\rangle\simeq \dfrac{2\cdot 10^{-27}\: (\mbox{cm}^3 /\mbox{s})}{\Omega_{DM}
h^2} \label{a22}
\end{equation}
For the present value \citep{wmap} of $\Omega_{DM} h^2 = 0.113$ we
obtain $\langle\sigma\upsilon\rangle\simeq 2\cdot 10^{-26}\:
(\mbox{cm}^3 /\mbox{s})$, i.e., a cross section typical for weak
interactions. Besides, the DMP must be massive for the dark matter
to be cold in the epoch when the relic radiation was generated.
These are the telling reasons to believe that the dark matter
consists of Weakly Interacting Massive Particles (WIMPs). They are
expected to annihilate into fermion-antifermion or gauge boson
pairs with a large fraction of quark-antiquark pairs. A WIMP pair
annihilation finally leads to $30-40$ photons generation (in the
fragmentation process, mainly from $\pi^0$ decays). A greater part
of the photons has energy in the range from 2 to 4~{GeV}
\citep{jungman, kazakov}.

Photon propagation at cosmological redshifts was extensively
investigated in \citet{rassey}. For the photons of energy
2-4~{GeV} the main channel of interaction with the baryonic matter
is pair production on atoms and ionized matter ($\gamma A\to A e^+
e^-$). As we shall see below, the Universe becomes transparent to
2 - 4~{GeV} photons remarkably after the recombination. The pair
production cross-section per one baryon averaged over the chemical
composition (mass fraction of hydrogen and helium are 75\% and
25\%, respectively) is
\begin{equation}
\aleph= 4.64\, \alpha r_e^2 \ln\left(\dfrac{513\mu}{\mu+825}\right) \label{a23}
\end{equation}
Here $\mu\equiv \dfrac{\varepsilon_\gamma}{m_e c^2}$, $r_e\equiv \dfrac{e^2}{m_e c^2}$. As we can
see, the cross-section depends only slightly on the energy of photon. $\aleph=1.63\cdot
10^{-26}\,\mbox{cm}^2$ if $\mu=4000$. Now we should substitute the obtained numerical values for
$\langle\sigma\upsilon\rangle$ and $\aleph$ into (\ref{a20}), but before we can improve the model.
To do this, we roughly estimate the number and the energy of photons generated by the process
$\gamma A\to A e^+ e^-$. It produces one electron and one positron: each of them carries away
approximately half of the energy of the primary photon. Subsequently, the positron annihilates with
an electron generating two photons. Hence, their energy is, on average, a quarter of the
primary one. Of course, these photons should have rather a wide distribution but we neglect it and
think of that one primary photon of energy $\varepsilon$ produces two secondary photons of energy
$\varepsilon/4$. Since the cross-section (\ref{a23}) only slightly depends on the photon energy,
we suppose that a secondary photon interacts with the matter with the same cross-section $\aleph$
as the primary one.

We consider the secondary photon scattering as pure absorption.
Indeed, the photons generated by them via the pair production have
energies $\sim 100$~{MeV}. In this energy range the Compton
scattering becomes the main process of interaction \citep{rassey}.
Its cross-section rapidly grows with the photon energy decreasing,
and low-energy photons are scattered time and again, quickly
loosing the energy. Thus, a scattering of a photon of energy $\le
400$~{MeV} is actually equivalent to its absorption.

The number of the secondary photons is symbolized by $N_2$. In order to obtain a differential
equation describing their propagation we can use the same procedure as in (\ref{b2}-\ref{a20}). The
relationship (\ref{b3}) acquires the form
\begin{equation}
dN_2 = 2 N \wp \, \dfrac{d\chi}{\chi^2 \sqrt{\chi}} -N_2 \wp \, \dfrac{d\chi}{\chi^2 \sqrt{\chi}}
\label{a24}
\end{equation}
The second term here describes the secondary photon absorption by the substance, the first one
represents their production via the above-mentioned mechanism (the multiplier $2$ takes into
account that one act of a primary photon absorption produces two secondary photons). The number of
primary photons $N$ is given by (\ref{a17}). As the boundary condition we should use the fact that
there were no secondary photons when the primary had been just generated
\begin{equation}
\left.\phantom{\int} N_2 \right|_{\chi=a} = 0 \label{a25}
\end{equation}
The solution of (\ref{a24})  satisfying the conditions is
\begin{eqnarray}
N_2 = N_{initial} \: \frac43 \wp \left(\dfrac{1}{a \sqrt{a}}-\dfrac{1}{\chi
\sqrt{\chi}}\right)\times \nonumber \\
\times \exp\left(\frac{2}{3} \frac{\wp}{\chi\sqrt{\chi}}-\frac{2}{3} \dfrac{\wp}{a\sqrt{a}}\right)
\label{a26}
\end{eqnarray}
In order to calculate the value for the observer we should substitute $\chi=a_0=1$. Since $a\ll 1$
\begin{equation}
N_2 = N_{initial} \: \frac43 \wp \dfrac{1}{a \sqrt{a}} \exp\left(-\frac{2}{3}
\dfrac{\wp}{a\sqrt{a}}\right) \label{a27}
\end{equation}
Here $N_{initial}$ is the initial number of primary photons. As in the case of the derivation of
 equation (\ref{a20}), we substitute (\ref{a12}) for $N_{initial}$
\begin{equation}
P_2=\dfrac{\langle\sigma\upsilon\rangle n_0^2 c}{6\pi H_0 \sqrt{\Omega_m}}\: \dfrac{\wp}{a^4}
\exp\left(-\frac{2}{3} \dfrac{\wp}{a\sqrt{a}}\right) dt \, do\, da\, dS \label{a28}
\end{equation}
We suppose that the energy of a secondary photon is a quarter of the energy of the primary
one. To account for this effect, $\dfrac{4 \varepsilon}{\beta}$ should be substituted for $a$
instead of (\ref{a8}). Finally, we obtain for the flux $Q_2$ of secondary photons:
\begin{equation}
Q_2=\dfrac{\langle\sigma\upsilon\rangle n_0^2 c \wp}{384 \pi H_0 \sqrt{\Omega_m}}\:
\dfrac{\beta^3}{\varepsilon^4} \exp\!\left(-\frac{\wp}{12} \, \dfrac{\beta
\sqrt{\beta}}{\varepsilon \sqrt{\varepsilon}}\right) \label{a29}
\end{equation}
The total signal is the sum $Q+Q_2$. If the annihilation generates photons with some
distribution $f$ (one act of annihilation produces $f(\beta)\, d\beta$ photons in the energy
interval $d\beta$), we should convolute the sum with the distribution, by analogy with (\ref{a21}):
\begin{equation}
\tilde Q(\varepsilon)=\int (Q(\varepsilon,\beta) + Q_2(\varepsilon,\beta))\cdot f(\beta)\, d\beta
\label{a30}
\end{equation}

\begin{figure}
 \resizebox{\hsize}{!}{\includegraphics{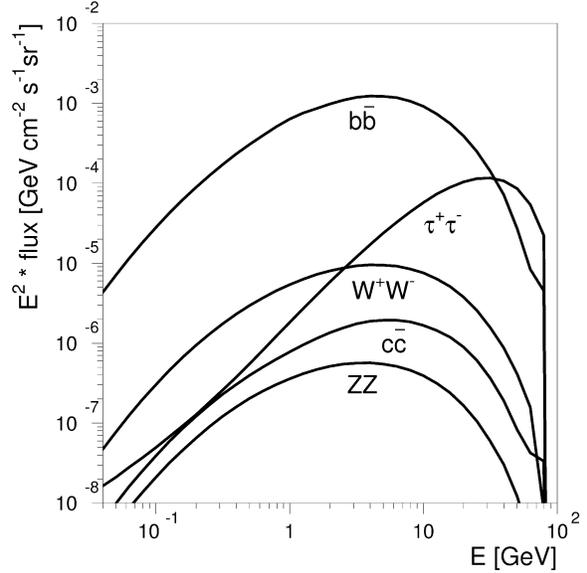}}
 \vspace{-4.5cm}
 \caption{The spectra of photons generated by various WIMPs annihilation channels, taken from
 \citet{kazakov} (the normalization is arbitrary, the WIMP mass is 100~{GeV}). The $b \bar b$ channel is most important
 \citep{jungman}.} \label{fig1}
\end{figure}
Figure~\ref{fig1} (taken from \citet{kazakov}) represents the photon spectra
caused by various WIMPs annihilation channels. In the interval $1 - 5$~{GeV}
the upper curve (which describes the main $b \bar b$ channel) is well
approximated if the distribution function $f(\beta)\propto
\beta^{-1}\exp(-0.15\beta)$. We shall use the following $f$ ($\beta$ is
expressed in GeVs):
\begin{equation}
f(\beta)= \begin{cases}
26.7\cdot\beta^{-1}\exp(-0.15\beta),&\text{$\beta\in(1 - 5)$ GeV}\\
0,&\text{$\beta\notin(1 - 5)$ GeV}
\end{cases}\label{a31}
\end{equation}
We normalized it considering that one act of annihilation generated on average 30 photons.

Strictly speaking, the annihilation cross-section and the relative
contribution of various channels depend on the energy of the DMPs
collision, which can be important, if the annihilation occurs, for
instance, near a black hole \citep{baushev2008bh}. However, in the
considered case the relative velocity of the DMPs is very small,
and we may neglect this effect.

\section{Comparison with observations and discussion}

In order to fulfil the calculations, the cosmological parameters
should be concretized. We use the following set (see
\citet{gorbrub} and references therein): $\Omega_\Lambda = 0.75$,
$\Omega_m = \Omega_{DM} + \Omega_b=0.25$, $\Omega_b = 0.042$ (of
course, $\Omega_{\Lambda}+\Omega_{DM}+\Omega_b=1$), the Hubble
constant $H_0=2.4\cdot 10^{-18} \mbox{s}^{-1}$, the relic
radiation temperature $2.725$~K, the baryon-photon ratio
$\eta\equiv n_b/n_{ph}=6.1 \cdot 10^{-10}$. We obtain the present
baryon concentration $n^b_0 = 2.5 \cdot 10^{-7}\, \mbox{cm}^{-3}$.
The DMP concentration, with DMP mass taken as $M_{DMP}=50$~{GeV},
is $n_0=2.5 \cdot 10^{-8}\, \mbox{cm}^{-3}$. Equation (\ref{a22})
gives $\langle\sigma\upsilon\rangle\simeq 2 \cdot 10^{-26}\:
(\mbox{cm}^3 /\mbox{s})$.

The influence of the relic annihilation on the ionization history of the
universe is negligible. Indeed, for the recombination epoch ($z\simeq 1200$) we
have:  $n^b \simeq n^b_0 z^3= 4.3\cdot 10^2 \, \mbox{cm}^{-3}$, $ n \simeq n_0
z^3 =43\, \mbox{cm}^{-3}$. The number of annihilations in a unit volume per
unit time is $\dfrac12 \langle\sigma\upsilon\rangle\: n^2 = 1.9\cdot 10^{-23}
{cm}^{-3} {s}^{-1}$. In the characteristic time (of the order of the hydrogen
ion recombination time at that epoch {435}~{years}~$\simeq 1.4\cdot 10^{10}$~s
\citep{gorbrub}) we have $2.6 \cdot 10^{-13}$ annihilations that produce $2.6
\cdot 10^{-2}$~{eV} of energy. A hydrogen atom ionization requires $\sim
14$~{eV}. So, even all the energy produced by the annihilation is enough to
ionize only $4\cdot 10^{-4}$~\% of atoms.

We take $\aleph=\aleph(2 \mbox{GeV}) =1.63 \cdot 10^{-26}\,\mbox{cm}^2$
(\ref{a23}). Then the constant $\wp$ is equal to $1.0 \cdot 10^{-4}$.
\begin{figure}
 \resizebox{\hsize}{!}{\includegraphics[angle=270]{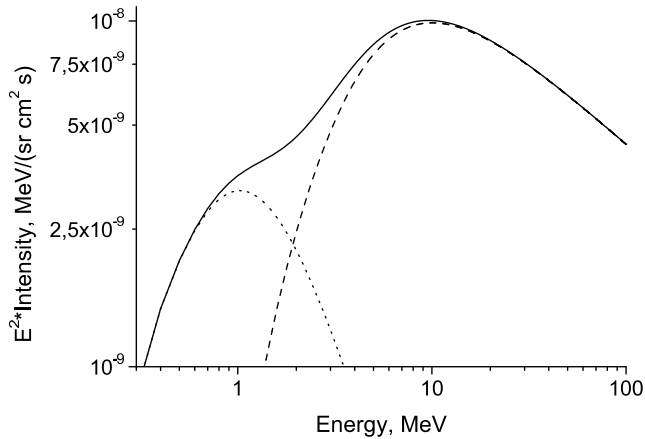}}
 \caption{The spectrum of the relic gamma-ray background, calculated according to (\ref{a30}),
(\ref{a31}), (\ref{a20}), and (\ref{a29}). Separate contributions of the primary and the
scattered photons are represented by the dashed and the dotted lines, respectively.} \label{fig2}
\end{figure}
The resulting spectrum of the relic gamma-ray background, obtained with the aid
of (\ref{a30}), (\ref{a31}), (\ref{a20}), and (\ref{a29}), is represented in
Figure~\ref{fig2}. Instead of the photon flux $\tilde Q$ we have plotted the
quantity $\varepsilon^2 * \mbox{flux}$ (that is $\varepsilon^2 \cdot \tilde
Q$), which is traditionally used in experimental data picturing. The primary
(\ref{a20}) and the secondary (\ref{a29}) photon separate contributions are
represented by the dashed and the dotted lines, respectively. The contribution
of secondary photons is small in these coordinates, even though their total
number is very large. Our main conclusions are that the spectrum grows up to
$\sim 8$~{MeV}, and the bulk of the signal lies in the range from 0.5 to
20~{MeV}. Characteristic redshift of the relic gamma-rays can be easily
calculated. According to (\ref{a20}), the quantity $\varepsilon^2 \cdot \tilde
Q$ has its maximum at $\frac{\varepsilon}{\beta} = a = (2 \wp)^{2/3}$.
According to (\ref{b1}), it corresponds to $z \simeq 300$.

The cosmic gamma-ray background reportedly (see, for instance, \citet{exp1,
ahn2005a, ahn2005b, rasera2006, strigari2005}) has a peculiarity in the energy
range 0.5 - 20~{MeV}. A ledge-like feature is visible in the extragalactic
gamma-ray spectrum (Fig.~3 in \citet{strong2004a}). The photon index here is
markedly distinct from those of the softer or harder parts of the spectrum
\citep{sreekumar1998, strong2004a, weidenspointner2000}, indicating its
different origin.

Moreover, this spectral band can be formed neither by too soft
emission of normal active galactic nuclei, nor by too hard
blazar-type AGNs contribution (see \citet{exp1} and references
therein). Attempts to consider the nuclear-decay gamma rays from
Type Ia supernovae as the source have not been successful: the
flux expected from the supernovae is several times weaker than the
observed \citep{ahn05, rasera2006, strigari2005}. It might be well
to point out that the precise determination of the excess
boundaries and intensity is model-dependent, and the literature
values vary considerably \citep{exp1, ahn2005a, ahn2005b,
rasera2006}. In any case, however, the excess becomes apparent
near 0.5~{MeV} and disappears at the energies $\gtrsim$~20~{MeV}.
One can see that the energy range of the feature corresponds
closely to the interval characteristic for the relic gamma
emission from the WIMPs annihilation. This coincidence looks
promising when it is considered that the WIMP is now one of the
most probable dark matter candidates.

At the same time, the relic gamma emission predicted by equations
(\ref{a30}), (\ref{a31}), (\ref{a20}), and (\ref{a29}), is
approximately five orders fainter than the observed feature (as we
can see in Figure~\ref{fig2}, the relic emission near the maximum
at 10~{MeV} has $\varepsilon^2 * \mbox{flux}\simeq 10^{-8}\,
\mbox{MeV} /(\mbox{sr}\, \mbox{cm}^2\, \mbox{s})$ while the total
extragalactic gamma-ray background at 10~{MeV} has $\varepsilon^2
* \mbox{flux}\simeq 2\cdot 10^{-3}\, \mbox{MeV} /(\mbox{sr}\,
\mbox{cm}^2\, \mbox{s})$ \citep{sreekumar1998, rasera2006}). This discrepancy
might result from inapplicability of the assumption of homogeneous dark matter
distribution. In fact it cannot be so. The modern structure of the Universe
appeared from some initial perturbations that had already existed, beyond any
doubt, in the epoch $z \sim 300$. According to WMAP measurements \citep{wmap},
in the recombination epoch $z \simeq 1100 \div 1400$ relative variations of the
baryonic matter density were of the order of $10^{-5}$. Dark matter
perturbations could be much more intensive, they were not suppressed by the
radiation pressure in the pre-recombination epoch. Moreover, they must have
been significantly stronger (not less than $10^{-3}$) to explain the modern
Universe structure \citep{gurevichzybin}. Since the recombination happened in
the matter-dominated epoch, the perturbations rapidly grew and at the moment $z
\sim 300$ could attain very big amplitude. The presence of density
inhomogeneities does not affect the spectrum of the annihilation signal but
increases its intensity. This effect is usually described by the quantity
$$
C\equiv \dfrac{<\rho^2>}{<\rho>^2}
$$
that appears as a multiplier in the expression for the intensity (see, for
instance, \citet{ahn2005a}). Of course, $C$ is a function of $z$.

This brings up two points: first, is it possible that the structure growth in
the early universe proceeded so fast that $C$ was as large as $10^5 - 10^6$ by
the moment $z=300$? Second, if an intensive structure formation took place at
some moment $\tilde z < 300$, the coefficient $C(z)$ could grow so rapidly that
it far outweighed the signal diminution owing to average density decrease. As a
result a hard tail or even a secondary hard maximum on the energy (2-4)/$\tilde
z$~{GeV} can appear in the spectrum of the relic emission, which is not
observed. Unfortunately, the theory of evolution of dark matter perturbations
is still far from accurate. They evolved from some primordial fluctuations
existed in the very early universe. While the universe was radiation-dominated
their growth was slow. The smallest perturbations were destroyed by
free-streaming (for the instance of neutralinos of the mass $\sim 100$~{GeV}
this limit is estimated as $10^{-12}-10^{-6} M_\odot$ \citep{10-12, 10-6}).
When the universe transits into the matter-dominated stage, the perturbations
start to grow rapidly, and eventually they become nonlinear and collapse. As
this takes place, the smallest clumps collapse the first (at a time we denote
as $\tilde z$) \citep{gurevichzybin}. The overwhelming majority of these small
clumps originated at that moment were subsequently destroyed by tidal
interaction with larger clumps originated later. But up to now it is the small
clumps formed at the epoch $\tilde z$ that makes the main contribution to the
annihilation rate of the dark matter since they are the densest
\citep{berezinsky2003, gurzybsir}. At the moment $\tilde z$ function $C(z)$
underwent a rapid increase from a value close to $1$ to a very big value. In
what follows creation of the larger structures was accompanied by smaller clump
destruction, and $C(z)$ changes more smoothly. In order to be allowed to
suggest that the extragalactic gamma-ray 0.5 - 20~{MeV} excess is related to
the neutralino annihilation we must assume that $\tilde z > 300$, i.e. the
first structures started to form before the universe became transparent for the
photons produced by the annihilation.

How realistic is such an assumption? Unfortunately, present estimations are
very vague. Even the minimal possible clump mass for the neutralino dark matter
is determined extremely uncertainly (from $10^{-12} M_\odot$ \citep{10-12} to
$10^{-6} M_\odot$ \citep{10-6}, to say nothing of the clumps density profile
and the moment when the fluctuations become nonlinear. Experimental data as
well as numerical simulations essentially cover the range of very large
structures $10^{15}-10^6 M_\odot$ (for instance, WMAP can observe only the
biggest perturbations with masses corresponding to a cluster of galaxies $\sim
10^{15} M_\odot$ \citep{wmap}). Properties of smaller clumps are usually
obtained by approximation \citep{ahn2005a}. However, in order to obtain any
parameters for the tiny clumps of mass $10^{-6}-10^{-12} M_\odot$ one has to
extend the results by $12$-$18$ orders. Another source of uncertainties is the
spectrum of primordial fluctuations. Usually it is deemed that it has flat
Harrison-Zeldovich shape. In this case the moment of the first intensive clump
creation is estimated as $\tilde z \simeq 80$, though some individual clumps
collapsed much earlier \citep{green2005}. If such a scenario was indeed
realized, the interpretation of 0.5 - 20~{MeV} excess as a result of neutralino
annihilation is out of the question: otherwise a strong maximum at the energy
4/$\tilde z$~{GeV}~$\simeq$~50~{MeV} would appear (which contradicts to the
observations).

On the other hand, let us suppose that the spectrum of primordial fluctuations is not exactly flat,
and the intensity of the fluctuations slightly builds up as their scale decreases. In the
matter-dominated stage the perturbations grow as $\delta\rho/\rho\propto t^{2/3}$ and $a(t)\propto
t^{2/3}$, therefore $\delta\rho/\rho\propto a$ \citep{gurzybsir}. If we assume that the small-scale
perturbations left the radiation-dominated stage with amplitudes more than expected from the
Harrison-Zeldovich spectrum, they collapsed earlier (in so doing we imply that the amplitudes of
large-scale perturbations are fixed in such a way as they reproduce the observed large-scale
structure of the Universe). Considering that the largest and the smallest clump mass scales differ
by more than $20$ orders, even a small tilt of the spectrum of primordial fluctuations with respect
to the Harrison-Zeldovich shape can be sufficient. A kindred scenario was considered, for instance,
by \citet{gurzybsir}. In the context of the model examined by the authors small clumps are
extremely dense and collapse just after the universe transition to the matter-dominated epoch
($\tilde z>1500$). It is worthy of note that the situation when $\tilde z
> 300$ can appear in much softer scenarios than those similar to
\citet{gurzybsir}.

Let us give a more specific form to the above reasoning. The spectrum of
primordial fluctuations is usually considered to have a power-law shape:
\begin{equation}
\label{c1}
 |\delta^{2}({\bf k})|\sim k^n
\end{equation}
The case when $n=1$ corresponds to the Harrison-Zeldovich spectrum. The moment
when a perturbation mode becomes nonlinear and collapses is determined by the
spectrum function \citep{gurzybsir}
\begin{equation}
\label{c2}
 \Gamma(k)\propto\dfrac{k^3 |\delta^{2}({\bf k})|}{1+(k/k_{eq})^4}
\end{equation}
(in \citet{gurzybsir} $\Gamma(k)$ is symbolized by $F(k)$). For all the
perturbations we consider $k\gg k_{eq}$, and
\begin{equation}
\label{c3}
 \Gamma(k)\propto\dfrac{|\delta^{2}({\bf k})|}{k}\propto k^{n-1}
\end{equation}
The moment $a^*=\frac{1}{1+z^*}$ when clumps collapse is
\begin{equation}
\label{c4}
 a^*\propto\dfrac{1}{\sqrt{\Gamma(k)}}\propto k^{-\frac{n-1}{2}}
\end{equation}
which correlates with the results of \citet{bullock}. In the case of Harrison-Zeldovich spectrum
$\Gamma(k)$ is flat; the structures of various scales appear almost simultaneously and have similar
densities. If $n>1$, smaller clumps appear much earlier and have much higher density (since the
universe density is much higher at the moment they collapse).

We can now construct a toy model of structure formation with the help of an
approach used in \citep{ahn2005a, bullock}. We take $|\delta^{2}({\bf k})|\sim
k^2$, i.e. $n=2$, and the minimal possible clump mass $M_{min}=10^{-7}
M_\odot$. Since the clamp mass relating to a perturbation mode is proportional
to $k^{-3}$, we obtain from (\ref{c4}):
\begin{equation}
\label{c41}
 a^*= \left(\dfrac{M}{M_{max}}\right)^{\frac{n-1}{6}} =
 \left(\dfrac{M}{M_{max}}\right)^{\frac{1}{6}}
\end{equation}
Here we have introduced the maximum perturbation mass $M_{max}$ that collapses at $a=1$. This
equation is consistent with \citet{bullock}, where it was accepted $M_{max}\simeq 1.5 \cdot 10^{13}
h^{-1} M_\odot$ ($h=0.7$). Then the first clumps in our model appear at $\tilde a=\frac{1}{2450}$,
i.e $\tilde z = 2450$. We can rewrite (\ref{c4}) as
\begin{equation}
\label{c8}
 a^*= \tilde a\cdot\left(\dfrac{M}{M_{min}}\right)^{\frac{1}{6}}
\end{equation}
In accordance with \citet{ahn2005a} we can represent the boost factor $C(z)$ as a product of three
multipliers:
\begin{equation}
\label{c5}
  C(z) =\Delta(z) \cdot F_{coll}(z) \cdot [C^{halo}]
\end{equation}
(see all the details of the model and the notation in \citep{ahn2005a,
bullock}). For a flat $\Lambda$CDM universe $\Delta(z)=(18\pi^2+82x-39
x^2)/(x+1)$, where $x=(\rho_m/\rho_{cr})-1$ and $\rho_m$, $\rho_{cr}$ are the
matter and the critical universe densities at given $z$.

The matter fraction collapsed into cosmological halos is:
\begin{equation}
\label{c6}
 F_{coll}(z) = \int_{M_{min}}^{M^*}\frac{dn}{dM} M dM/\rho_0
\end{equation}
Hereafter $M^*$ is the maximum having collapsed at given $z$. Theoretical model
\citep{berezinsky2003} and numerical simulations \citep{diemand} give for the
differential number density of clumps in the comoving frame of reference
$\frac{dn}{dM}\propto M^{-2}$. Substituting it in (\ref{c6}) and using
(\ref{c8}), we obtain:
\begin{equation}
\label{c7}
 F_{coll}(z) = \frac{F^0_{coll}}{46.8}
 \ln\left(\dfrac{M^*}{M_{min}}\right)=\frac{F^0_{coll}}{7.8}
 \ln\left(\dfrac{a}{\tilde a}\right)
\end{equation}
where $F^0_{coll}\equiv F_{coll}(a=1)$. We will adopt the value obtained by
\citet{ahn2005a} $F^0_{coll}\simeq 0.8$.

The factor $[C^{halo}]=\int C^{halo} M \frac{dn}{dM} dM/ \int dM
\frac{dn}{dM}M$ represents the "halo clumping". If we adopt the
Navarro-Frenk-White clump profile than
\begin{equation}
\label{c9}
 C^{halo}=\frac{c_{vir}^{3}(1-1/(1+c_{vir})^3)}{9(\ln(1+c_{vir})-c_{vir}/(1+c_{vir}))^2}
\end{equation}
In order to describe the halo concentration parameter $c_{vir}$ evolution we
will use an equation from \citet{ahn2005a, bullock} with $K=8$, $F=0.01$
\begin{equation}
\label{c10}
 c_{vir}(M,a)=K a \left(F \dfrac{M}{M^*}\right)^{-1/6}= K \dfrac{a}{\tilde a}
\left( \dfrac{M_{min}}{F M}\right)^{1/6}
\end{equation}
This formula was obtained as a fit of N-body simulations and is valid only for
a limited range of $M$ and $z$ covered by them. We have to consider a much
wider mass and red shift range, and (\ref{c10}) gives too large $c_{vir}$ for
the smallest clumps (for instance, for the minimal mass clumps at the present
epoch $c_{vir}\simeq 20000$). Such a huge value seems unlikely and indicates
that equation (\ref{c10}) should be corrected. Following \citet{ahn2005a}, we
assume that for any chosen clump mass the concentration parameter rises only up
to $c_{vir}=100$, and after remains constant.
\begin{figure}
 \resizebox{\hsize}{!}{\includegraphics[angle=270]{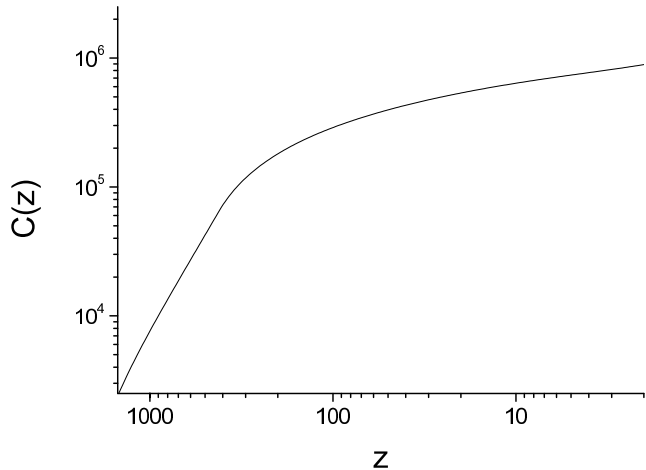}}
 \caption{The boost factor evolution with $z$.}
 \label{fig3}
\end{figure}

Now the toy model is defined. The boost factor $C(z)$ evolution curve predicted by it is
represented in Figure~\ref{fig3}. We can make two principal conclusions. First of all, $\tilde z =
2450$ and $C(z=300)\simeq 1.5\cdot 10^{5}$ in the model considered, i.e. the first structures
appear very early ($\tilde z \gg 300$), and at the moment $z=300$ the boost factor is large enough
to explain the discrepancy between the observed and the predicted signal intensities. Second, the
boost factor $C(z)$ grows respectively slowly after $z=300$. From the moment when the universe
became transparent for the annihilation photons up to the present moment it increases only on an
order, while the signal without regard for the clumpiness rapidly falls as $z^{2.5}$ owing to the
dark matter density decrease (\ref{a12}). It means that almost all the signal appears at the moment
$z\simeq 300$ and the resulting spectrum has neither a hard tail nor a secondary harder maximum. In
closing we remark that the foregoing structure formation scheme is no more than a toy model
presented here only to illustrate that the structures in the universe could appear early, and the
interpretation of the excess as a result of WIMP annihilation is principally possible.

The dark matter annihilation has already been invoked to explain the 0.5 -
20~{MeV} excess \citep{ahn2005a, ahn2005b, rasera2006}. Since the photon
spectrum is relatively soft, the authors introduced a low-mass dark matter
candidate ($M_{DMP} < 100$~{MeV}). Its annihilation cross-section is sizable to
provide the observed signal $\langle\sigma\upsilon\rangle\simeq 2.5\cdot
10^{-26}\: (\mbox{cm}^3 /\mbox{s})$ \citep{ahn2005a} that is at least no less
than the typical weak interaction cross-section at this energy scale. Such a
low-mass dark matter candidate with such a significant cross-section is now
ruled out (by the accelerator experiments) in ordinary schemes like MSSM, but
there is an interesting possibility to introduce it in more sophisticated
scenarios \citep{boehm1, boehm2}.

In our case, we can manage with usual heavy WIMP candidates. As we could see,
the relic gamma-ray signal redshift is $z \sim 300$. Originally, the photons
had the energy $1 \div 5$~{GeV} and they were produced by the annihilation of
ordinary WIMP particles like the lightest neutralinos. On the other hand, if
our interpretation is true, it counts in favour of a relatively light WIMP
($M_{DMP} \sim 100$~{GeV}). If $M_{DMP} \gg 100$~{GeV}, the typical energy of
the producing photons is higher, and the feature in the spectrum must be harder
than the observed one.

Thus, the critical point for the 0.5 - 20~{MeV} excess interpretation as a
result of relic neutralino annihilation is determination of the moment $\tilde
z$ when the first structures in the universe appeared. If the structures had
appeared before the universe became transparent for the annihilation products
(i.e. $\tilde z > 300$), then there are strong arguments to believe that the
excess 0.5 - 20~{MeV} is created by the relic WIMPs annihilation. Above all,
the characteristic energies of the spectra agree. Besides, the WIMP is now one
of the most probable dark matter candidates, and the coincidence does not look
random. The discrepancy of the predicted and observed signal intensities can be
naturally explained by the nonuniform structure of the dark matter. If the
first clumps appeared later ($\tilde z < 300$), the excess undoubtedly could
not be produced by neutralino annihilation. Further progress in the Universe
structure formation understanding will be able to shed light on this problem.

\section{Acknowledgements} This work was supported by the RFBR (Russian Foundation for Basic
Research, Grant 08-02-00856).

\label{lastpage}
\end{document}